\pdfoutput=1
\documentclass[11pt]{article}

\PassOptionsToPackage{table,dvipsnames}{xcolor}
\usepackage[final]{acl}
\tcbuselibrary{skins,breakable}
\usepackage{times}
\usepackage{latexsym}
\usepackage[T1]{fontenc}
\usepackage[utf8]{inputenc}
\usepackage{microtype}
\usepackage{float}
\usepackage{amsmath,amssymb}
\usepackage{graphicx}
\usepackage{booktabs}
\usepackage{multirow}
\usepackage{array}
\usepackage{enumitem}
\usepackage{placeins}
\newcommand{\benchmark}{\textsc{EntLORE}}
\newcommand{\methodname}{\benchmark}
\newcommand{\taskname}{\emph{implicit enterprise process question answering}}
\newcommand{\capabilityname}{\emph{latent organizational reasoning}}
\newcommand{\papertitle}{\benchmark: A Graph-Grounded Benchmark for Latent Organizational Reasoning in Enterprise Question Answering}

\newcommand{\numdocuments}{2{,}341}
\newcommand{\numdoctypes}{three}
\newcommand{\numquestions}{907}
\newcommand{\graphstatistics}{1{,}153 entities and 3{,}784 typed relations}
\newcommand{\numconfigurations}{56}
\newcommand{\nummodels}{8}

\title{\papertitle}
\author{
\textbf{Akrin Zheng\textsuperscript{*}},
\textbf{Alexander Wu\textsuperscript{*}},
\textbf{Alaia Liu\textsuperscript{*,\dag}}
\\
ScitiX.ai
\\
\small{\textbf{\textsuperscript{*}These authors contributed equally to this work.}}
\\
\small{\textbf{\textsuperscript{\dag}Correspondence:}
\href{mailto:alaia@scitix.ai}{alaia@scitix.ai}}
}

\begin{document}
\maketitle

\begin{abstract}
Enterprise question answering is framed as retrieving internal
documents and generating grounded answers. Routine enterprise records,
however, are work by-products in which required organizational relations
remain implicit across heterogeneous sources. Existing benchmarks provide realistic multi-source evidence, but often materialize a
predefined answer path and therefore test the composition of stated facts
rather than recovery of a target relation absent from the corpus. We call the
latter capability \capabilityname{}.

We introduce \textbf{\methodname{}}, a graph-grounded benchmark construction framework
that reconstructs an audited enterprise world from routine documents,
authoritative organizational tables, and operational records. Versioned
organizational conventions certify derived relations in a truth graph,
enabling complete golden answers and proof certificates. The aligned
anonymized release exposes only the document corpus while withholding private
structure and target relations. \textbf{\benchmark{}} contains \textbf{\numdocuments{}} documents
from \numdoctypes{} source types and \numquestions{} questions spanning
explicit lookup, cross-source composition, and latent organizational
reasoning, evaluated across \numconfigurations{} model and access
configurations. Structuring the released world as an induced entity graph or
navigable knowledge base gives the strongest deployable results. Yet supplying
gold documents still leaves \textbf{30.4\%} of latent questions unanswered, versus
\textbf{12.6\%} and \textbf{6.2\%} for explicit and compositional questions. Enterprise QA
therefore depends not only on document recall, but also on whether implicit
organizational relations become usable. The benchmark, data, and code are
publicly available at \url{https://github.com/scitix/entlore}.
\end{abstract}
\section{Introduction}
\begin{figure}[H]
\centering
\includegraphics[width=0.95\linewidth]{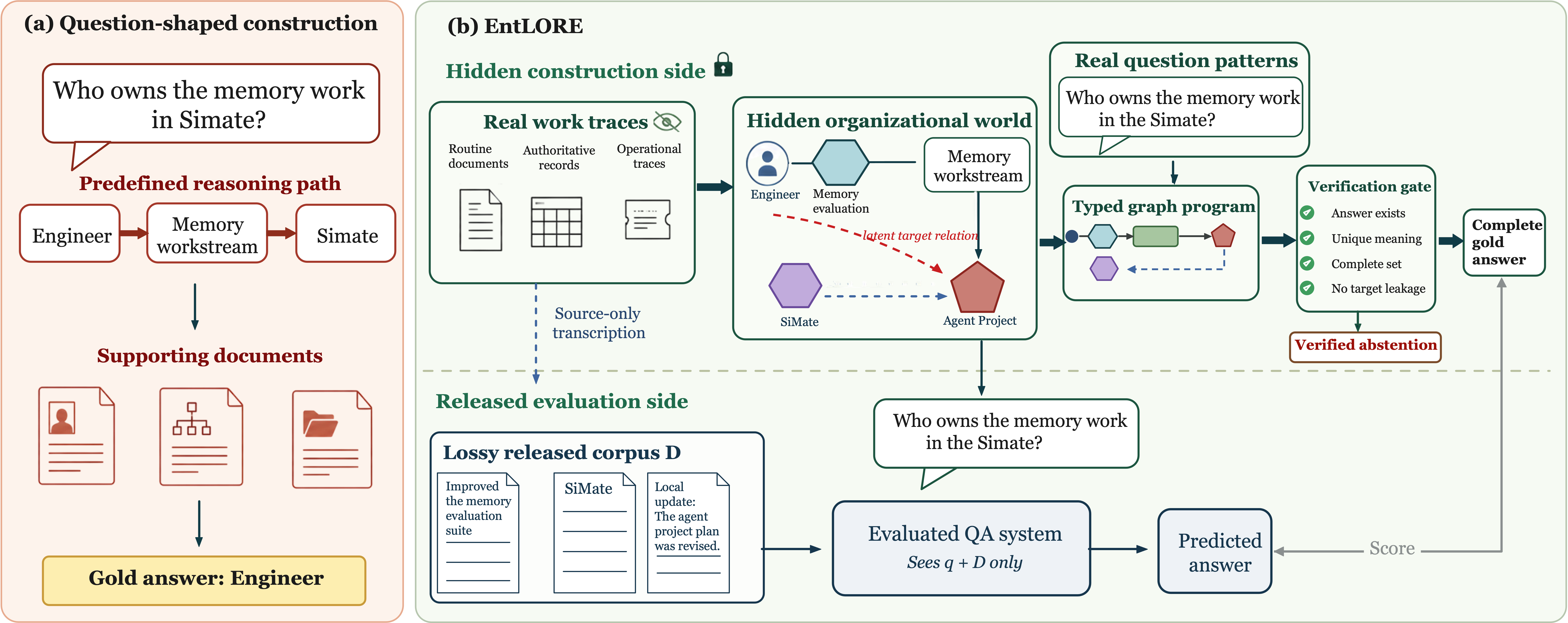}
\caption{Query-shaped construction versus \benchmark{}. Controlled benchmarks
materialize documents along a predefined reasoning path (left). \benchmark{}
reconstructs an audited enterprise world and releases an aligned anonymized
corpus; graph programs derive answers and proofs while target relations remain
withheld (right).}
\label{fig:teaser}
\end{figure}

Enterprise question answering is typically formulated as a retrieval problem:
given a user query, the system identifies relevant internal documents and
generates an answer grounded in them. In practice, however, many enterprise
questions cannot be answered by retrieving an explicitly stated fact.
Internal documents are usually produced as by-products of day-to-day work
rather than as self-contained descriptions written for future question
answering.


Recent enterprise QA benchmarks have substantially improved multi-source
evaluation through workflow-guided synthesis, project-coherent corpora, and
executable event representations
\citep{choubey2025herb,sun2026enterpriserag,cohen2025wixqa,qobench2026}.
These approaches provide controlled evidence composition and reliable
supervision. In many such constructions, however, evidence is organized around
a predefined answer path: each required step is materialized as explicit
document content, often distributed across multiple sources. The resulting
challenge is to retrieve and compose these stated facts, rather than to recover
a target relation absent from the corpus. Routine enterprise sources provide
no such explicit chain. The project scope of an activity, the correspondence
between an internal alias and a broader initiative, or responsibility for a
workstream may remain implicit across heterogeneous records. We call the
ability to recover and reason over these relations \capabilityname.

This setting also creates a benchmark-construction problem. A realistic
enterprise information need is not necessarily answerable from the released
knowledge sources. A structural premise may remain private, the visible
records may cover only part of the answer, or the question may admit several
organizational interpretations. Directly generating questions from selected
documents avoids this uncertainty by tailoring each question to visible
evidence, but it cannot establish whether realistic enterprise questions
have complete and uniquely defined answers.

Rather than shaping documents around a predefined reasoning path, our
graph-grounded construction framework first reconstructs an audited enterprise
world from heterogeneous private sources and then compiles realistic enterprise
question families into executable graph programs (Figure~\ref{fig:teaser}).
The graph serves as an intermediate semantic layer between
questions and fragmented sources. It separates audited base facts from weak
associations, certifies organization-specific derived relations, filters out
unsupported or ambiguous candidates, computes complete golden answers, and
links every premise to a released evidence unit. Question wording is realized
with controlled personas and audited for organizational plausibility; the
graph determines whether the question is answerable and what constitutes its
correct answer.

The resulting benchmark is an aligned, anonymized projection of the source
enterprise world rather than a corpus synthesized around individual questions.
The release preserves the forms, source-derived content, and cross-document
information structure of routine enterprise records, while consistently
replacing identity-bearing details across documents and questions. Its audited
questions reflect realistic enterprise information needs and remain aligned
with the released world.

Using this framework, we construct \benchmark{}, containing \numdocuments{}
documents from \numdoctypes{} enterprise source types and \numquestions{}
questions. We define a \numconfigurations{}-configuration evaluation matrix
spanning different models and access paradigms. Access conditions that
materialize organizational structure over the released world lead the matrix,
plain lexical access remains a strong baseline, and flat dense retrieval and
agentic pipelines lose the largest share of their attainable performance on
latent organizational questions.

Our contributions are as follows:
\begin{itemize}
  \item We formulate \taskname{} and identify \capabilityname{} as a central
  enterprise QA capability: systems must combine heterogeneous grounded facts
  to recover organization-specific relations before answering many routine
  questions.
  \item We construct an audited benchmark truth graph from real enterprise
  documents, authoritative organizational tables, and operational records,
  separating native, metadata, heuristic, and certified derived relations.
  \item We introduce \methodname{}, a graph-grounded construction framework
  that jointly determines answerability, complete golden answers, proof
  dependencies, and release-grounding status without exposing instance-level
  gold relations to evaluated systems.
  \item We construct \benchmark{} and evaluate models and knowledge-access
  systems across explicit lookup, cross-source composition, latent
  organizational reasoning, evidence acquisition, answer generation, and
  release fidelity.
\end{itemize}

\section{Related Work}

\paragraph{Enterprise and document-grounded QA.}
Realistic domain QA has moved from curated passages toward large,
heterogeneous collections. TechQA uses authentic technical-support questions
over a corporate corpus; doc2dial and MultiDoc2Dial ground information-seeking
interactions in one or multiple service documents; and EKRAG benchmarks
retrieval and answer generation over corporate releases, blogs, and reports
\citep{castelli2020techqa,feng2020doc2dial,feng2021multidoc2dial,yu2025ekrag}.
WixQA combines real user queries with expert-written answers that can span
multiple knowledge-base articles, while EnterpriseRAG-Bench synthesizes
coherent company corpora with question-aware evidence structures
\citep{cohen2025wixqa,sun2026enterpriserag}. Financial-document benchmarks
require joint reasoning over prose, tables, and numerical programs, including
FinQA, TAT-QA, and MultiHiertt, while ARQA and MuDABench emphasize
deterministic verification and aggregation over annual-report collections
\citep{chen2021finqa,zhu2021tatqa,zhao2022multihiertt,wang2026arqa,li2026mudabench}.
These resources improve domain and corpus realism, but their answers are
generally grounded in facts explicitly available in released documents,
knowledge bases, or deterministic table and database programs. They therefore
primarily evaluate retrieving and composing stated information rather than
recovering an organization-specific target relation absent from the corpus.

\paragraph{Multi-hop and executable benchmark construction.}
HotpotQA, 2WikiMultiHopQA, IIRC, and MuSiQue distribute explicit supporting
facts across documents and annotate facts, paths, answerability, or connected
reasoning
\citep{yang2018hotpotqa,ho2020twowiki,ferguson2020iirc,trivedi2022musique}.
KILT standardizes a shared corpus and provenance, KQA Pro compiles questions
from a knowledge base into executable programs, and ProofWriter evaluates rule
application with explicit proofs
\citep{petroni2021kilt,cao2022kqapro,tafjord2021proofwriter}. QO-Bench comes
closest to our executable perspective by computing deterministic answers over
typed event tuples recovered from corporate text, but targets database-style
operations over event attributes rather than organization-specific derived
relations \citep{qobench2026}. Document multi-hop benchmarks thus distribute
explicit reasoning steps, while executable benchmarks operate over
materialized knowledge, rules, or typed records. Our setting instead
reconstructs and certifies organizational relations from routine records, then
withholds the target relations from the released evidence.

\paragraph{Retrieval-augmented and graph-based access.}
DPR, RAG, and FiD establish retrieve-and-read or retrieve-and-generate
architectures; IRCoT and ReAct interleave retrieval with reasoning or actions
\citep{karpukhin2020dpr,lewis2020rag,izacard2021fid,trivedi2023ircot,yao2022react}.
RAPTOR organizes long documents hierarchically, Self-RAG adaptively retrieves
and critiques, and HippoRAG, G-Retriever, and GraphRAG exploit graph structure;
CRAG broadens evaluation across web and knowledge-graph access
\citep{sarthi2024raptor,asai2024selfrag,gutierrez2024hipporag,he2024gretriever,edge2024graphrag,yang2024crag}.
These methods differ in how they retrieve, organize, or navigate released
evidence and therefore form complementary evaluation subjects for
\benchmark{}. They may induce useful structure, but do not by themselves
provide benchmark-side certification of which unstated organizational
relations are valid, complete, and grounded in released premises. \benchmark{}
uses a private audited truth graph for that certification while withholding
its target relations from evaluated systems. Across these lines, prior work
improves corpus realism, reasoning control, or knowledge access; \benchmark{}
complements them by evaluating whether systems can recover an audited
organizational relation that no released document states.

\section{The \methodname{} Framework}
\label{sec:method}

\begin{figure}[t]
\centering
\includegraphics[width=0.98\textwidth]{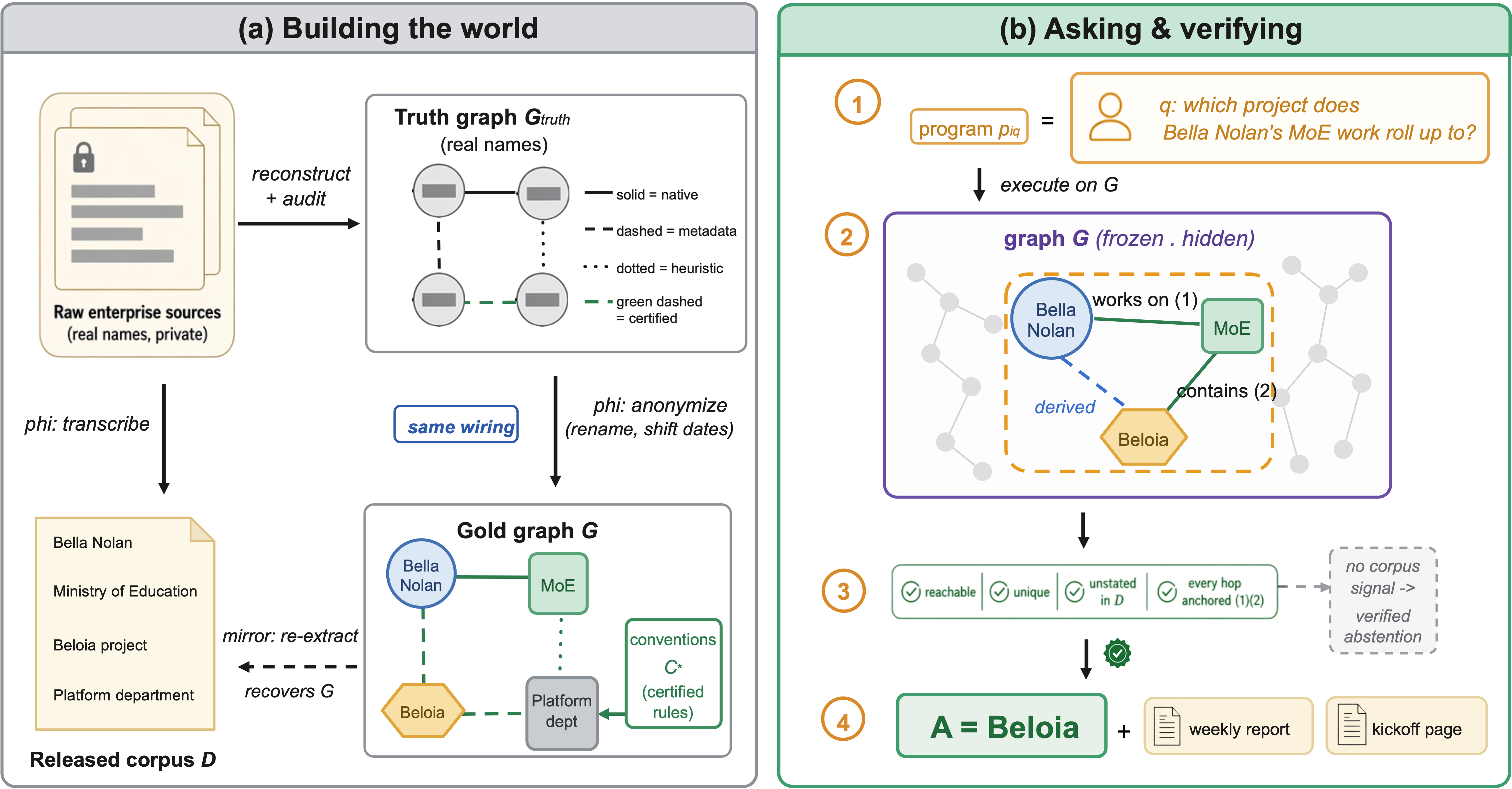}
\caption{
The \methodname{} pipeline.
Heterogeneous enterprise sources are reconstructed into a raw graph.
Audited organizational conventions add certified inference edges to form the
truth graph. A shared anonymization map produces the released document corpus;
private organizational records, certification rules, and the truth graph
remain hidden. Typed graph programs compute answers and verify their
derivations, completeness, uniqueness, and target absence from released
documents.
}
\label{fig:pipeline}
\end{figure}

\methodname{} constructs an enterprise QA benchmark in four stages: it
reconstructs a provenance-bearing raw graph from routine documents,
authoritative records, and operational traces; certifies authorized
organization-specific inferences to form a truth graph; projects this world
into an anonymized release; and compiles each question into an executable graph
program that computes its golden answer and certifies its derivation,
completeness, and unique denotation. Routine sources expose only partial
observations, so the framework
separates the private graph that certifies answers from the released evidence
world available to systems. Figure~\ref{fig:pipeline} shows the pipeline.

\subsection{Audited Truth Graph and Released Evidence World}
\label{sec:world}

We reconstruct a real-named raw graph $G_{\mathrm{raw}}$ over people,
projects, modules, campaigns, systems, and customers from routine documents,
authoritative organizational records, and operational traces. Each edge
retains source provenance. Native relations are grounded in documents or
traces; metadata relations come from authoritative records; and heuristic
associations support candidate discovery and decoy construction but never
independently establish gold truth.

A versioned convention library produces candidate organizational inference
edges. A retained edge records its generating rule, supporting dependencies,
and human authorization verdict. The resulting truth graph is

\[
G_{\mathrm{truth}}
=
\operatorname{audit}(G_{\mathrm{raw}})
\cup
\mathcal{C}^{*}(G_{\mathrm{raw}}),
\]

where $\mathcal{C}^{*}$ denotes certified rule applications rather than an
unrestricted closure. A shared anonymization map $\phi$ produces the released
document corpus $D_{\mathrm{release}}$ by consistently renaming private
entities and aliases and shifting dates while preserving their relative
organization. Evaluated systems access only $D_{\mathrm{release}}$; private
organizational records, $G_{\mathrm{truth}}$, certification rules, target
edges, and item-level proof certificates remain hidden.

\subsection{Organizational Conventions}
\label{sec:graph}

Many enterprise questions target relations that no released source states
explicitly. We encode the organizational knowledge needed to certify them as
private typed conventions that are not released to evaluated systems. For
example, containment transfer attributes module-level activity to its parent
project:

\[
\begin{aligned}
&\textsf{works\_on}(p,m) \circ
\textsf{contains}(\mathit{proj},m)\\
&\qquad\Rightarrow
\textsf{attributed\_to}(p,\mathit{proj}).
\end{aligned}
\]

Additional convention families, their typed preconditions, and applicability
conditions are given in Appendix~E. Each
certified edge
retains its rule version and dependency chain.

\subsection{Privacy-Preserving Corpus Projection}
\label{sec:projection}

Each released document is an anonymized transcription of exactly one raw
document. The projection preserves genre, local organization, information
density, and natural omissions, but introduces no facts from organizational
tables, other documents, or certified inference edges. Thus, abbreviated
names, omitted project scope, and reliance on prior context remain properties
of the released corpus rather than generation artifacts. Shared anonymization
preserves cross-source identity and temporal structure; detailed privacy and
fidelity controls appear in Appendix~D.

\subsection{Question Construction and Capability Decomposition}
\label{sec:questions}

Each question instance is generated by a typed graph operator that samples an
eligible region of $G_{\mathrm{truth}}$, executes a query, and records its typed
answer, proof dependencies, and associated released evidence units. A language
model then realizes the question in a concise employee-like form without
receiving the golden answer. Graph operators determine question semantics;
language models determine only surface wording.

We group questions by how the target answer relation is represented in the
released evidence:

\begin{itemize}
    \item \textbf{L1 --- stated lookup.}
    The answer is explicitly stated in one released document.

    \item \textbf{L2 --- stated composition.}
    The answer composes facts explicitly stated across multiple released
    sources.

    \item \textbf{L3 --- latent organizational reasoning.}
    The target relation appears in no released source. It is certified only in
    $G_{\mathrm{truth}}$ and must be recovered from observations distributed
    across the released document corpus.
\end{itemize}

We use these levels as a capability decomposition. L1 evaluates single-source
evidence acquisition and answer extraction, while L2 evaluates cross-source
composition when all required facts are explicitly stated. Together they
provide diagnostic baselines for the prerequisite evidence-use capabilities.
L3 builds on these prerequisites while withholding the target relation from
the released world; producing a grounded answer therefore additionally
requires \capabilityname{} from organizational structure implicit across the
released documents.

For example, one report may state that an engineer worked on a module, while
other released documents place that module within a project's work context.
No released source directly attributes the engineer to the project; the truth
graph certifies this attribution through the applicable containment-transfer
convention.

\subsection{Verification and Golden Answers}
\label{sec:verify}

Every question is verified before release. The verifier checks that the graph
program returns an answer with valid dependencies and certified derivations,
that set-valued answers are complete under the relevant closure, and that the
program has a unique denotation. For an L3 item, the verifier additionally
requires a proof containing a certified organizational derivation, and a
release-wide scan verifies that the target relation is absent under every
released entity alias.

Graph execution establishes correctness and completeness relative to the
audited enterprise world. Human semantic review separately verifies that the
natural-language question expresses the intended graph program and does not
admit a materially different interpretation.

Directly stated values are anchored to source spans; relations, sets, and
aggregates are computed by typed graph programs. Automated release gates
check provenance, replay, target absence, and privacy, followed by human
review. Candidates requiring an indispensable private-only premise are routed
to the verified-abstention bank. Full gates are listed in Appendix~C; scoring
rules appear in the Experiments section.

\section{Experiments}
\label{sec:experiments}

\begin{table}[!t]
\centering
\small
\setlength{\tabcolsep}{0.9mm}
\renewcommand{\arraystretch}{0.9}
\resizebox{\textwidth}{!}{%
\begin{tabular}{l*{21}{c}}
\toprule
\multirow{2}{*}{Model}
& \multicolumn{3}{c}{Closed-book}
& \multicolumn{3}{c}{BM25}
& \multicolumn{3}{c}{RAG}
& \multicolumn{3}{c}{Agentic}
& \multicolumn{3}{c}{LLM Wiki}
& \multicolumn{3}{c}{GraphRAG}
& \multicolumn{3}{c}{Oracle} \\
\cmidrule(lr){2-4}
\cmidrule(lr){5-7}
\cmidrule(lr){8-10}
\cmidrule(lr){11-13}
\cmidrule(lr){14-16}
\cmidrule(lr){17-19}
\cmidrule(lr){20-22}
& L1 & L2 & L3
& L1 & L2 & L3
& L1 & L2 & L3
& L1 & L2 & L3
& L1 & L2 & L3
& L1 & L2 & L3
& L1 & L2 & L3 \\
\midrule
GPT-5.4           & 1.1 & 0.0 & 8.3 & \textbf{63.0} & 51.8 & 34.9 & 47.9 & 34.3 & 19.1 & 49.1 & 40.5 & 17.5 & 62.3 & \textbf{52.5} & \textbf{35.5} & 56.7 & 51.5 & 34.1 & 88.9 & 95.2 & 69.5 \\
GPT-5.4-mini      & 0.0 & 0.0 & 7.7 & \textbf{61.9} & 45.1 & \textbf{33.0} & 44.3 & 27.9 & 18.4 & 42.5 & 30.4 & 14.8 & 55.2 & \textbf{47.1} & 26.5 & 52.2 & 32.9 & 28.8 & 87.5 & 84.9 & 62.1 \\
Claude-S 4.6      & 0.9 & 0.0 & 8.5 & 64.0 & \textbf{51.7} & 36.4 & 46.8 & 32.1 & 21.2 & 54.7 & 35.6 & 14.9 & \textbf{66.7} & 51.5 & 34.6 & 65.8 & 43.7 & \textbf{38.3} & 88.5 & 94.5 & 76.3 \\
Qwen3.5-397B      & 0.4 & 0.0 & 7.7 & 64.9 & \textbf{53.0} & 36.2 & 48.4 & 33.2 & 19.5 & 46.0 & 35.0 & 9.5 & 59.4 & 49.3 & 20.4 & \textbf{65.6} & 46.6 & \textbf{39.6} & 85.2 & 95.9 & 62.5 \\
GLM-5.2           & 0.5 & 0.0 & 7.7 & 60.9 & 51.2 & 30.9 & 44.1 & 32.1 & 17.4 & 51.0 & 35.9 & 12.3 & 67.0 & \textbf{53.2} & 30.8 & \textbf{68.5} & 46.5 & \textbf{43.1} & 89.2 & 95.3 & 70.9 \\
Kimi-K2.6         & 0.4 & 0.0 & 8.1 & \textbf{63.8} & \textbf{55.2} & \textbf{35.1} & 46.4 & 37.0 & 19.5 & 46.5 & 31.5 & 9.4 & 57.2 & 49.7 & 14.0 & 63.4 & 40.9 & 27.6 & 88.2 & 94.9 & 70.6 \\
DS-V4-Pro         & 0.7 & 0.0 & 7.7 & 61.2 & 52.6 & 33.8 & 46.1 & 33.3 & 17.7 & 48.4 & 39.1 & 14.3 & 64.9 & \textbf{56.9} & 29.6 & \textbf{66.2} & 48.8 & \textbf{38.6} & 86.4 & 95.2 & 73.5 \\
DS-V4-Flash       & 1.1 & 0.0 & 7.7 & 59.3 & 50.8 & 31.8 & 45.0 & 30.2 & 18.4 & 49.6 & 36.4 & 15.1 & 61.4 & \textbf{59.6} & 30.6 & \textbf{67.6} & 51.7 & \textbf{39.8} & 85.6 & 94.7 & 71.3 \\
\midrule
\textit{Mean}     & 0.6 & 0.0 & 7.9 & 62.4 & 51.4 & 34.0 & 46.1 & 32.5 & 18.9 & 48.5 & 35.6 & 13.5 & 61.7 & \textbf{52.5} & 27.7 & \textbf{63.2} & 45.3 & \textbf{36.2} & 87.4 & 93.8 & 69.6 \\
\bottomrule
\end{tabular}
}
\caption{Overall answer accuracy (\%) by level (L1/L2/L3) across \nummodels{}
models and seven access conditions. LLM Wiki is an LLM-compiled offline wiki
over the released corpus; $\Omega$ is a gold-document agentic reference that
exposes the gold document set through the same tool loop without supplying the
target relation. Boldface marks the best deployable condition per model and
level; closed-book and $\Omega$ are excluded from that comparison.}
\label{tab:overall_perf}
\end{table}

\begin{figure}[t]
\centering
\includegraphics[width=0.98\textwidth]{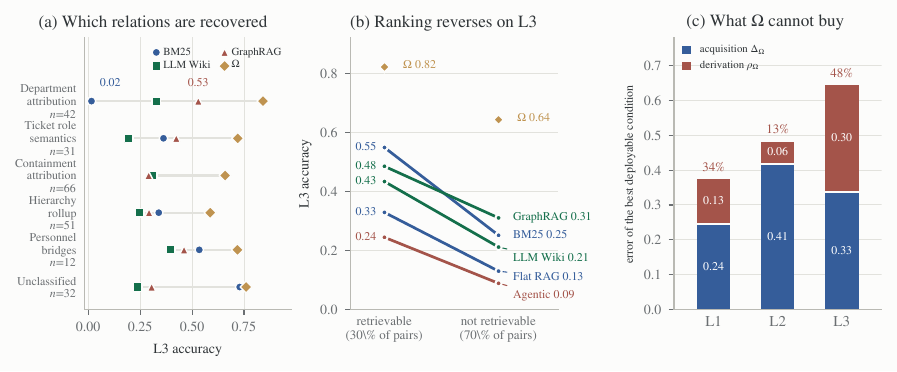}
\caption{Performance across L3 relation families, flat-retriever
evidence assembly, and gold-document residuals. (a) L3 accuracy by certified relation
family, with BM25, LLM Wiki, GraphRAG, and $\Omega$ as the four markers;
families are ordered by the GraphRAG-minus-BM25 gap. Flat RAG and agentic
retrieval are omitted for legibility; they are the bottom tier in every
family. \emph{Unclassified} collects the L3 operators outside the five
certified families. (b) L3 items grouped by whether either evaluated
single-pass flat retriever assembled the complete annotated gold-document set,
with $\Omega$ marked per group. Because the grouping is defined by BM25 and
dense-retrieval outcomes, it is used descriptively; $\Omega$ shows that the
groups also differ in difficulty. (c) The remaining error of the best
deployable condition at each
level (GraphRAG at L1 and L3, the LLM wiki at L2), split algebraically into
the oracle gap $\Delta_{\Omega}$ and residual $\rho_{\Omega}=1-\Omega$, with
the residual share printed above each bar.}
\label{fig:analysis}
\end{figure}

The benchmark levels separate prerequisite evidence use from organizational
relation recovery. We therefore first examine whether systems can handle
stated evidence, then ask what changes when the target relation is absent from
every released document. We ask four questions.
\textbf{RQ1:} How well do models and access methods handle stated lookup and
cross-source composition?
\textbf{RQ2:} When the target relation is absent from every released document,
how do access paradigms compare, and where do their differences concentrate?
\textbf{RQ3:} What remains unresolved when the gold documents are supplied?
\textbf{RQ4:} Does the anonymized release preserve the comparative behavior of
the private source world?
Each question is answered by one analysis below.

\subsection{Experimental Setup}

\paragraph{Benchmark and task.}
\benchmark{} contains \numquestions{} questions (469 L1, 204 L2, 234 L3) over
\numdocuments{} released documents and an organizational graph of
\graphstatistics{}. A system receives a question and the released world
$D_{\mathrm{release}}$ under its access paradigm; no system receives private
organizational records, certification rules, the truth graph, or instance-level
gold relations. An L3 answer must therefore be recovered from the released
document corpus alone.

\paragraph{Systems.}
We compare five deployable access paradigms, a closed-book floor, and a
gold-document oracle reference:
\begin{itemize}
  \item \textbf{Closed-book}: the model receives only the question.
  \item \textbf{BM25}: a lexical retriever ranks the released evidence units;
  its top-ranked units are supplied in a single answer-generation call. No
  structure is built.
  \item \textbf{Flat RAG}: the same units are embedded into one dense retrieval
  index and supplied in a single generation call \citep{lewis2020rag}. No
  structure is built.
  \item \textbf{Agentic retrieval}: the model uses the same dense index,
  embeddings, evidence units, and top-$10$ retriever as Flat RAG, but may call
  search or fetch tools over at most 30 LLM iterations \citep{yao2022react}.
  Each item is additionally capped at 400{,}000 cumulative input and output
  tokens and 900 seconds. Successful early termination requires an explicit
  answer submission; exhausting any budget without submission is scored as an
  incorrect answer.
  \item \textbf{LLM-Wiki}: following the llm-wiki paradigm~\cite{karpathy2026llmwiki}, we compile the corpus offline into a navigable wiki using an implementation based on Google's Open Knowledge Format~\cite{google2026okf}, without access to benchmark questions or gold relations. An LLM assigns free-form concept types, descriptions, and tags. Page bodies preserve the released source documents verbatim, cross-links are generated via exact title matching, and generated text is restricted to metadata and directory-level navigation summaries.

  \item \textbf{Corpus-induced GraphRAG}: an entity graph, its typed relations,
  and community reports are induced offline from the same corpus; the system
  never consumes the benchmark truth graph \citep{edge2024graphrag}.
  \item \textbf{$\Omega$---Gold-document agentic reference}: the model is
  given the gold document paths and reads them through the same agentic tool
  loop. This removes document-search recall errors but retains navigation,
  interaction, and answer-generation effects. $\Omega$ never supplies the
  target relation; for L3 it remains absent from every document.
\end{itemize}
All deployable systems are constructed from the same released document corpus
and receive no benchmark questions or gold relations during indexing. They
intentionally differ in indexing, offline compilation, query-time interaction,
and computational cost. Flat RAG and agentic retrieval share the same dense
index and retrieval backend, isolating the effect of iterative interaction.
Agentic retrieval and the LLM wiki use the same agent-loop framework but expose
different knowledge substrates, isolating the effect of offline compilation.
GraphRAG is a broader system-level structured-access alternative. We report
\textsc{oracle-agentic} as $\Omega$; the remaining oracle variants, model and
protocol details, and per-type scoring rules are in Appendix~G.

\paragraph{Models and scoring.}
Models are GPT-5.4, GPT-5.4-mini, Claude-Sonnet-4.6,
Qwen3.5-397B-A17B, GLM-5.2, Kimi-K2.6, DeepSeek-V4-Pro, and
DeepSeek-V4-Flash. Deterministic answer types are scored programmatically;
free-form answers are reduced to atomic claims and scored by a single fixed
judge, blind to system identity (judge-replication results in Appendix~G).

\paragraph{Metrics.}
Using $\Omega$ as the gold-document reference, we define for system $s$ and
level $\ell$:
\begingroup\small
\[
\underbrace{1-\operatorname{Acc}(m,s,\ell)}_{\text{total}}
=
\underbrace{\Omega(m,\ell)-\operatorname{Acc}(m,s,\ell)}_{\Delta_{\Omega}:\ \text{access gap}}
\;+\;
\underbrace{1-\Omega(m,\ell)}_{\rho_{\Omega}:\ \text{residual}} .
\]
\endgroup
These quantities are diagnostic rather than causal. The oracle gap
$\Delta_{\Omega}$ includes differences in finding, organizing, presenting, and
navigating evidence. The residual $\rho_{\Omega}=1-\Omega$ contains every error
that remains under the gold-document agentic condition. On L3 it is an upper
bound on failures of organizational relation recovery, not a pure derivation
measure.

\subsection{RQ1: Stated-Evidence Performance}

Table~\ref{tab:overall_perf} reports standard-condition performance for all
eight answer models and all seven access conditions. All five deployable
conditions are built from the same released documents without gold
supervision, but intentionally differ in indexing, offline construction,
query-time interaction, and compute.

On L1, GraphRAG, BM25, and the LLM wiki average $63.2\%$, $62.4\%$, and $61.7\%$;
on L2, the LLM wiki and BM25 lead at $52.5\%$ and $51.4\%$. Flat RAG and agentic
retrieval remain well below this group on both levels. Across all three
levels, BM25, GraphRAG, and the LLM wiki form an upper tier at $52.6\%$, $52.2\%$,
and $50.9\%$ question-weighted accuracy, compared with $36.5\%$ for agentic
retrieval and $36.0\%$ for Flat RAG. The ranking is nevertheless
model-contingent: Kimi-K2.6 alone performs best with BM25 at all three levels.
Closed-book accuracy is $0.6\%/0.0\%/7.9\%$, so no condition is carried by
parametric memory. These results establish the stated-evidence capabilities
that L3 builds on.

\subsection{RQ2: Access Organization for Latent-Relation Questions}

RQ2 compares complete access paradigms on L3 and examines how their performance
differences vary across relation families and evidence-assembly conditions.

Averaged across the eight answer models, GraphRAG obtains the highest L3
accuracy at 36.2\%, followed by BM25 at 34.0\%, the LLM wiki at 27.7\%, Flat RAG
at 18.9\%, and agentic retrieval at 13.5\%. GraphRAG's mean advantage over BM25
is modest and model-dependent: it leads for five models, whereas BM25 leads
for three. Thus, the induced graph is the strongest average L3 condition in
our matrix, but structured access does not uniformly dominate lexical
retrieval.

The differences are highly heterogeneous across certified relation families
(Figure~\ref{fig:analysis}a). Department attribution produces the largest
separation: GraphRAG reaches 53\%, compared with 2\% for BM25, on 42 questions
for which the gold-document reference reaches 84\%. In contrast, containment
attribution and hierarchy rollup remain difficult for every deployable
condition, with no method exceeding 34\% against gold-document references of
66\% and 59\%. The largest system separations therefore concentrate in
particular organizational relations rather than appearing uniformly across
L3.

As a secondary diagnostic, Figure~\ref{fig:analysis}b groups the 1{,}872
model-question pairs from the eight answer models and 234 L3 items by whether
either evaluated single-pass flat retriever assembled the complete annotated
gold-document set. Because this grouping is defined by BM25 and dense-retrieval
outcomes, we use it descriptively to localize performance differences rather
than as a general measure of retrievability. On the 560 pairs (30\%) for which
the gold set is assembled, BM25 leads at 54.9\%, compared with 48.5\% for
GraphRAG and 43.3\% for the LLM wiki. On the remaining 1{,}312 pairs (70\%),
GraphRAG leads at 31.0\%, followed by BM25 at 25.1\% and the LLM wiki at 21.1\%.
The gold-document reference also falls from 82\% to 64\%, showing that the
latter group is harder overall. The contrasting orderings localize GraphRAG's
relative strength in the group where the evaluated flat retrievers do not
assemble the complete annotated evidence set.

Within the shared dense substrate, agentic retrieval trails single-pass Flat
RAG both overall on L3 (13.5\% versus 18.9\%) and in the group where the evaluated
flat retrievers do not assemble the complete gold set (8.8\% versus 12.9\%).
In this dense-backed implementation, iterative interaction therefore does not
improve single-pass retrieval. With the gold document paths exposed through
the same loop, $\Omega$ reaches 69.6\% on L3, showing that the loop can use
supplied evidence substantially better than it acquires evidence under the
evaluated dense backend.

Among the conditions without materialized structure, BM25 exceeds Flat RAG by
16.2/18.9/15.1 points on L1/L2/L3 and agentic retrieval by 13.9/15.9/20.5
points on the same corpus. On items where both flat retrievers assemble the
gold set, the two score 75.7\% and 72.1\%, and this difference contributes only
8\% of their overall gap; the subset that only BM25 assembles contributes
71\%. This pattern associates most of the observed gap with document assembly
rather than answer performance after both systems deliver the annotated gold
set. The concentration of the gap is consistent with lexical matching
benefiting from sparse organizational anchors---project aliases, module names,
ticket terms, and role labels.

Finally, among zero-scored L3 answers, 53\% of BM25 outputs explicitly state
that the requested fact is not present, compared with 6\% for GraphRAG. The
rates show that response calibration and abstention behavior are another
dimension along which the systems differ. Overall, GraphRAG is strongest on
average, with its largest separation from BM25 concentrated in selected
relation families and in cases where the evaluated flat retrievers do not
assemble the complete gold set. BM25 nevertheless remains competitive, and
the comparison concerns complete access systems rather than individual
GraphRAG components.

\subsection{RQ3: What Remains with Gold Documents}
\label{sec:omega}

\paragraph{Gold documents close the L1 and L2 gap but leave a third of L3
unanswered.}
Figure~\ref{fig:analysis}c splits the remaining error of the best deployable
condition into the oracle gap $\Delta_{\Omega}$ and gold-evidence residual
$\rho_{\Omega}=1-\Omega$. With the gold document paths supplied through the
agentic interface, 12.6\% of L1 and 6.2\% of L2 remain unanswered, compared
with 30.4\% of L3. The L3 residual is 4.9 times the L2 residual and accounts
for 47.7\% of the remaining error of the best deployable condition. Because
$\Omega$ retains navigation, interaction, answer extraction, and prompt
effects, this residual is not a pure derivation estimate. On L3, where the
target relation is absent from every supplied document, it is an upper bound
on failures of organizational relation recovery. Appendix~F reports a
consistent but limited post-hoc check on matched anchor entities.

\subsection{RQ4: Release Fidelity}

The anonymised release must preserve the comparative behaviour of the private
source world, or the benchmark measures its own projection. Running the same
answer model as the main matrix on the 176 aligned items whose wording
reverse-renders cleanly, the coarse structure transfers: the three oracle
configurations occupy the top three positions in both worlds and the deployable
conditions stay below them, with mean $|\Delta|=0.125$. The fine ordering among
the deployable conditions does not, and the rank statistic says so---$\rho=0.71$
over eight configurations does not reach significance (exact permutation
$p=0.06$). We therefore read the release as validated for group-level comparison
and unproven for fine ranking. The deviation is concentrated in BM25 ($-0.28$)
and GraphRAG ($-0.19$) while Flat RAG and agentic retrieval are essentially
unchanged, consistent with lexical sensitivity to a corpus that is both
anonymised and translated. Appendix~A gives the per-configuration figure and
the extended interpretation.

\FloatBarrier
\section{Conclusion}

Enterprise QA requires more than retrieving stated facts: routine questions
may depend on organizational relations that remain implicit across
heterogeneous records. \methodname{} makes this setting evaluable by
reconstructing an audited truth graph, certifying organization-specific
derivations, and releasing an aligned anonymized document corpus while
retaining executable answers and proofs privately. The resulting \benchmark{}
separates explicit lookup and cross-source composition from latent
organizational reasoning without exposing target relations to evaluated
systems.

Across the same corpus, an induced entity graph and a navigable knowledge base
give the strongest deployable results, while plain lexical retrieval
outperforms flat dense and agentic retrieval. More importantly, supplying gold
document paths still leaves $30.4\%$ of latent organizational questions
unanswered. Although this residual is not a pure measure of derivation
failure, removing document-search errors does not remove the latent-relation
challenge. Release-fidelity analysis supports broad, group-level comparisons
while leaving fine-grained rankings tentative. Together, these findings suggest
that enterprise QA systems should treat the organization behind the documents
as a first-class inference object. Future systems should recover typed relations
across sources, preserve provenance through multi-step reasoning, and abstain
when an organizational inference is not warranted. By keeping benchmark truth
private but its premises auditable, \benchmark{} provides a testbed for measuring
such progress without reducing it to document recall alone.




\bibliography{references}

\clearpage
\appendix
\setcounter{secnumdepth}{2}

\section{Extended Release-Fidelity Analysis}
\label{app:release-fidelity}

\begin{figure}[H]
\centering
\includegraphics[width=0.92\columnwidth]{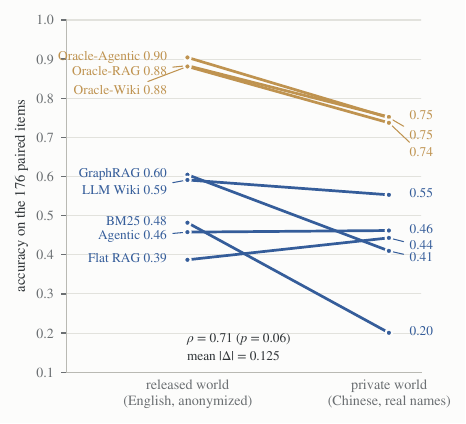}
\caption{\textbf{Private-to-release fidelity}, one line per access configuration
over the 176 paired items. Crossings are rank changes and the overall
displacement is the absolute shift; the three oracle configurations stay together
at the top of both worlds while the deployable ordering reshuffles.}
\label{fig:release-fidelity}
\end{figure}

Figure~\ref{fig:release-fidelity} shows the per-configuration detail.
We hypothesize that the BM25 discrepancy reflects lexical sensitivity to the
release projection. Chinese tokenization, identifier segmentation, and the
standardization of anonymized aliases can substantially alter lexical term
overlap without adding or removing an underlying organizational fact. The
result therefore does not imply that the release adds semantic information;
rather, it shows that lexical retrieval is especially sensitive to the textual
surface on which the same organizational information is expressed. We
accordingly interpret the release as preserving comparative evaluation
behavior rather than as reproducing private-world scores exactly.

Two asymmetries bound the audit. First, the private side reuses the indices
built once over the private corpus, whereas the released side uses the
per-model index builds of the main matrix; the comparison therefore holds the
index fixed on one side and not the other, and the retrieval-dependent
conditions inherit that difference. Second, the private corpus is the original
Chinese text while the released corpus is anonymized and translated, so the
BM25 deviation combines anonymization with a language change and should not be
read as an anonymization effect alone. The audit covers the 176 of 258 paired
items whose wording reverse-renders without a residual pseudonym; the remaining
82 are excluded because a compound pseudonym cannot be inverted cleanly, which
is a property of the rendering rather than of any evaluated system.

\section{Certified Relation Families}
\label{app:families}

The initial release focuses on relation families with explicit typed premises
and auditable business semantics:
\begin{itemize}
  \item \textbf{Containment attribution}: document-attested activity on a
  module combined with document-grounded module--project context.
  \item \textbf{Hierarchy rollup}: certified aggregation from modules to
  projects or campaigns when every child relation is release-grounded.
  \item \textbf{Personnel bridges}: people connecting two projects through
  separately grounded activities.
  \item \textbf{Ticket role semantics}: submitter, handler, assignee, owner,
  and affected party remain distinct typed roles.
  \item \textbf{Department attribution}: included only when the
  project--department premise has predicate-level document support.
\end{itemize}
Aggregation, footprint, handover, and time-window questions are retained as
secondary families when their base sets and time attributes are certified.
Private-only structural premises produce abstentions. Weak co-occurrence and
negative-space assumptions never define positive gold.

\section{Release Gate Pipeline}
\label{app:gates}

Gates run in order, and every rejected candidate is logged with its terminal
gate and reason:
\begin{itemize}
  \item \textbf{Version lock}: bind the item to pinned source, graph, corpus,
  schema, operator, rule, and realizer revisions.
  \item \textbf{Source-class legality}: every indispensable premise of an
  answerable item must be grounded in a released document; private metadata
  and heuristic associations may not independently establish answerability.
  \item \textbf{Rule certification}: verify rule identifier, version,
  applicability conditions, typed conclusion, and authorization verdict.
  \item \textbf{Predicate-level premise grounding}: resolve every premise to
  a released paragraph; endpoint co-occurrence is insufficient.
  \item \textbf{Public/private separation}: route any item with a required
  private-only premise to verified abstention.
  \item \textbf{Target absence}: confirm that an L3 target relation is absent
  from every released source.
  \item \textbf{Exact coverage}: compute the minimum set of released evidence
  units covering the proof.
  \item \textbf{Uniqueness and closure}: certify unique denotation or complete
  set closure under the audited graph scope.
  \item \textbf{Raw faithfulness}: trace every released document atom to its
  private source.
  \item \textbf{Privacy screening}: scan released strings and structures for
  source identifiers and high-risk rare details.
  \item \textbf{Question hygiene}: remove answer values, proof-path hints, and
  semantically ambiguous wording.
  \item \textbf{Raw-to-release replay}: verify identical typed program outputs
  before and after anonymization.
  \item \textbf{Bank-level controls}: remove near duplicates and enforce
  family, source, and answer-type balance.
\end{itemize}

Candidates are routed to answerable L1/L2/L3 when all proof premises are
release-grounded, to verified abstention when at least one required premise is
private-only, and to rejection when the target, proof, or question semantics is
invalid.

\section{Privacy and Identity Ledgers}
\label{app:privacy}

Persons, projects, customers, and internal systems are renamed under a shared
mapping. Alias families map isomorphically so that source alias-resolution
difficulty is preserved. Dates shift by a fixed whole-week offset to retain
reporting cadence. Released documents use stable anonymous identifiers. A
private source ledger records raw paths, hashes, per-atom support, metadata
origins, identity mappings, and reverse links from source records to questions.
The release process includes lexical leakage scanning, rare-event review, and
structural sensitivity checks. These checks provide release controls rather
than a general guarantee against all possible re-identification.

\section{Private Convention Library and Proof Certificates}
\label{app:conventions}

Every private convention entry contains:
\begin{itemize}
  \item a rule name and version;
  \item typed premise and conclusion schemas;
  \item applicability conditions and exclusions;
  \item the source class permitted for each premise;
  \item examples using generic entities only.
\end{itemize}
The initial library includes containment transfer, hierarchy rollup, selected
genre anchoring, and ticket role semantics. A default-scope rule is included
only after its exceptions---cross-team support, reviews, meetings, research,
and reported team-level progress---are explicitly encoded and audited.

Each certified derived relation carries a proof certificate
\[
(r,\mathrm{rule\_id},\mathrm{rule\_version},
E_{\mathrm{premise}},E_{\mathrm{release}},v),
\]
where $v$ is the human authorization verdict. A necessity certificate records
whether every valid derivation route requires at least one organizational
convention.

\section{Post-hoc Matched Triples}
\label{app:matched-triples}

\begin{figure}[t]
\centering
\includegraphics[width=0.98\columnwidth]{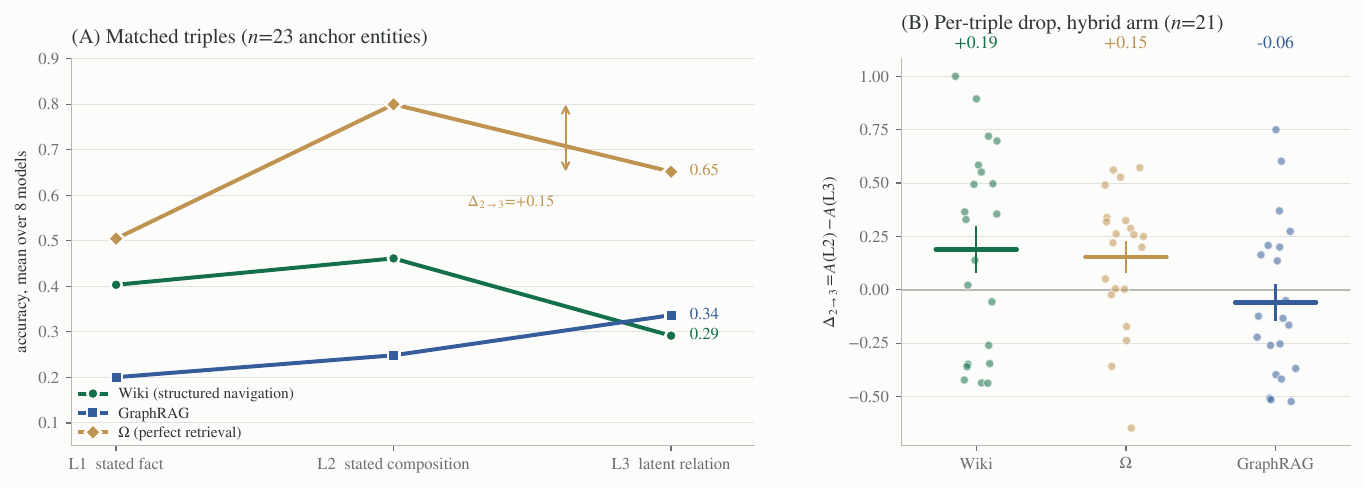}
\caption{\textbf{Post-hoc matched triples} ($n{=}23$ anchor entities; accuracy is
the per-item mean over the eight answer models). (a) Level slopes under the LLM
wiki, corpus-induced GraphRAG, and the gold-document agentic reference $\Omega$.
(b) The same $\Delta_{2\rightarrow3}$ per triple on the hybrid arm
($n{=}21$); bars are the mean, whiskers one standard error, and a positive value
means accuracy falls once the relation becomes latent.}
\label{fig:matched-triples}
\end{figure}

The three levels are separate question populations rather than a paired
construction. We therefore use this post-hoc analysis only as a consistency
check on the aggregate pattern, not as a controlled cross-level comparison. We
identify anchor entities that carry a question at all three levels and whose
L1 and L3 items share at least one released evidence document, after filtering
abbreviation noise. Twenty-three such triples exist
(Figure~\ref{fig:matched-triples}), 21 of them hybrid-source (the target
relation requires organizational structure) and only 2 text-grounded, so the
text-versus-hybrid contrast is underpowered and we do not draw it.

The check agrees with the oracle analysis in the main paper. Holding the anchor
entity and the evidence region fixed, accuracy falls from stated composition to
latent derivation by $\Delta_{2\rightarrow3}={+}0.19$ per triple on the hybrid
arm under the LLM wiki. Running the same triples under $\Omega$ reproduces the
drop with gold document paths supplied through the same agentic loop
($0.80 \rightarrow 0.65$), so it is not explained by document-search recall
alone. GraphRAG again
inverts it, rising from $0.25$ to $0.34$: materializing the organizational
relation supplies what the latent level withholds.

Two caveats bound this appendix. The triples are matched by anchor entity, not by
base premise, so an L1 and an L3 item in the same triple need not rest on the
same underlying fact; and $n{=}23$ is small. A designed premise-matched arm, in
which one base premise is escalated across the three levels, would replace this
check with a controlled one and is the natural next construction step.

\FloatBarrier
\begingroup
\small
\section{Experimental Details}
\label{app:experimental-details}

\paragraph{Benchmark statistics.}
The 907-item released bank contains 469 L1, 204 L2, and 234 L3 questions. Its
62 typed graph operators anchor questions to 1{,}150 distinct gold documents;
an item requires 2.53 gold documents on average and an L1 item requires 1.26.
The covered families include explicit lookup, cross-document composition,
latent project attribution, organizational aggregation, and incident roles.

\paragraph{Access implementations.}
BM25 ranks released evidence units lexically and supplies the highest-ranked
units in one answer-generation call. Flat RAG embeds the same units in one
dense index and likewise uses a single generation call. Agentic retrieval uses
that same dense index, embeddings, evidence units, and top-$10$ retriever. It
may invoke search or page-fetch tools over at most 30 LLM iterations; search
and fetch calls are choices within the iteration budget rather than separately
counted actions, and one iteration may contain multiple tool-use blocks. An
explicit answer submission is the only successful early termination. Each
item is also capped at 400{,}000 cumulative input and output tokens and 900
seconds of wall time; exceeding any cap without submission produces an error
string scored as an incorrect answer. When accumulated tool history exceeds
240{,}000 characters, earlier results are folded into a summary.

The LLM wiki compiles the corpus offline into a navigable collection with
free-form concept types, descriptions, and tags. Page bodies are lossless
copies of released source documents. Cross-links are generated through exact
title matching, while generated text is limited to metadata and
directory-level navigation summaries. The pipeline is fully automatic and
receives no benchmark questions, truth-graph edges, or gold relations.
GraphRAG induces an entity graph, typed relations, and community reports
offline from the same released corpus. Flat RAG and agentic retrieval share
the retrieval backend; agentic retrieval and the LLM wiki share the agent-loop
framework but expose different knowledge substrates.

\paragraph{Oracle variants.}
Four oracle forms vary only how the gold evidence is presented: canonical
paragraphs (\textsc{lc-oracle}), full documents in one context
(\textsc{oracle-rag}), document paths read page by page under a tool loop
(\textsc{oracle-agentic}), and the same documents routed through the compiled
knowledge base (\textsc{oracle-okf}). They diagnose paragraph localization,
document recall, navigation, and index-chain quality and pair with their
deployable counterparts. The main paper reports \textsc{oracle-agentic} as
$\Omega$: it exposes the gold document paths but retains the agentic navigation
and interaction layer. No oracle supplies the target relation.

\paragraph{Models and protocol.}
The evaluation covers eight models from the GPT-5.4, Claude, Qwen, GLM, Kimi,
and DeepSeek families. Exact versions and fixed context, output, reasoning,
and tool-use budgets are recorded in the versioned evaluation manifest.
All access and generation settings were fixed before the main evaluation; no
development-time hyperparameter search was conducted. Retrieval and
compilation run once per released world. Single-turn configurations run once;
agentic systems are additionally repeated on a stratified subset to measure
trajectory variance. The model-provider APIs do not expose a common,
reproducible seed control, so exact model and decoding settings are retained in
the manifest instead. Question-level bootstrap intervals quantify uncertainty,
and infrastructure failures are reported separately from model errors.

\paragraph{Code, data, and configuration availability.}
The Code and Data Supplement contains the complete released corpus and
question bank; preprocessing and benchmark-construction code; implementations
of the evaluated access conditions; prompts, scoring, analysis, and figure
code; the versioned evaluation manifest; and machine-readable outputs needed
to reproduce the reported aggregates. Upon publication, these artifacts will
be released under the MIT License.

\paragraph{Reasoning-trace setting and the empty-completion problem.}
In a first pass with each model's reasoning trace enabled, the five
open-weight models returned an empty completion on part of the closed-book,
BM25, and Flat RAG conditions, and the scorer counted every empty string as an
incorrect answer. The effect was large enough to distort a row rather than jitter
it: Qwen3.5-397B-A17B alone produced 1{,}228 empty completions, which depressed
its BM25 accuracy by roughly 19 points on L1 and 21 on L2. We therefore rerun
those three conditions in full, for all five open-weight models, with the
reasoning trace disabled, and the reported table uses the rerun for exactly
those cells. Every other cell---the agentic, LLM Wiki, GraphRAG, and $\Omega$
columns, and all three rows of GPT-5.4, GPT-5.4-mini, and Claude-Sonnet-4.6,
none of which produced empty completions---comes from the reasoning-enabled
pass. The mixture is therefore per condition and per model rather than global,
and it is the reason the closed-book, BM25, and Flat RAG columns should not be
read as a reasoning-enabled measurement for the open-weight models.

\paragraph{Score distribution at L3.}
L3 scoring is close to all-or-nothing rather than graded: under GraphRAG,
$49\%$ of L3 items score exactly zero and $23\%$ score exactly one, and
partial credit is almost entirely confined to set-valued answers. An overall
accuracy of $0.36$ at L3 therefore does not describe a system that is partly
right on most items; it describes one that is completely wrong on about half
and completely right on about a quarter.

\paragraph{A conflict between the oracle instruction and the latent level.}
The oracle prompt instructs the model to use only what it has read and not to
assert facts absent from the supplied pages. At L1 and L2 that instruction is
harmless, because the answer is stated in those pages. At L3 it is in tension
with the task, whose answer is by construction stated nowhere. The tension is
measurable but bounded: among L3 items the oracle scores zero on, $19.5\%$ of the
answers contain an explicit statement that the fact is not present, against
$4.0\%$ at L1 and $3.3\%$ at L2. Roughly a fifth of the L3 gold-evidence residual is
therefore attributable to instruction-following rather than to inability to
derive, and the residual should be read as an upper bound on organizational
relation-recovery failure.

The tension is total for one answer type. Eight L3 items ask for an integer count
over a latently derived set (\textsc{project\_headcount}); every model scores zero
on them under every condition, and the oracle recovers one item out of sixty-four
model-item pairs. Inspection shows the failure is not arithmetic: models enumerate
the contributing people correctly and then decline to state a total because no
document states one. One of these items also carries ambiguous wording that
several models read as asking about a population rather than a project roster.
These eight items measure the interaction between the instruction and the level
rather than the capability, and should be reworded or excluded before the bank is
frozen.

\paragraph{Answer judging.}
Free-form and atomic-claim scoring uses a single fixed judge, Claude Opus 4.8,
applied identically to every model and access condition; the judge never sees
which system produced an answer. Deterministic answer types (entity, set,
count, ordered) are scored programmatically and never reach the judge.
Replications with additional judge models are reported in the extended judging
analysis; they leave the reported orderings and conclusions unchanged.

\paragraph{Type-specific scoring.}
Entity answers use alias-normalized exact match. Set answers use precision,
recall, F$_1$, and exact-set accuracy. Integer counts require exact match;
tolerance applies only to explicitly continuous quantities. Ordered answers
use exact sequence accuracy and Kendall's $\tau$. Free-form answers are reduced
to typed atomic claims and scored by entailment against the graph answer and
proof, with unsupported additions counted as errors. Evidence precision and
recall use paragraph and metadata-record anchors.

\endgroup

\end{document}